\documentclass{article}
\usepackage{graphicx}
\usepackage{amsfonts}
\usepackage{amsmath}

\begin{document}

\title{Simple model of bouncing ball dynamics: displacement of the table
assumed as quadratic function of time}
\author{Andrzej Okninski$^{1}$, Bogus\l aw Radziszewski$^{2}$ \\
%EndAName
Physics Division$^{1}$, Department of Mechatronics and \\
Mechanical Engineering$^{2}$, \\
Politechnika Swietokrzyska, Al. 1000-lecia PP 7,\\
25-314 Kielce, Poland}
\maketitle

\begin{abstract}
Nonlinear dynamics of a bouncing ball moving in gravitational field and
colliding with a moving limiter is considered. Displacement of the limiter is a 
quadratic function of time. Several dynamical modes, such
as fixed points, $2$ - cycles and chaotic bands are studied analytically and
numerically. It is shown that chaotic bands appear due to homoclinic
structures created from unstable $2$ - cycles in a corner-type bifurcation.
\end{abstract}

\section{Introduction}

Vibro-impacting systems belong to a very interesting and important class of
nonsmooth and nonlinear dynamical systems \cite%
{diBernardo2008,Luo2006,Awrejcewicz2003,Filippov1988}\ with important
technological applications \cite%
{Stronge2000,Mehta1994,Knudsen1992,Wiercigroch2008}. Dynamics of such
systems can be extremely complicated due to velocity discontinuity arising
upon impacts. A very characteristic feature of such systems is the presence
of nonstandard bifurcations such as border-collisions and grazing impacts
which often lead to complex chaotic motions.

The Poincar\'{e} map, describing evolution from an impact to the next
impact, is a natural tool to study vibro-impacting systems. The main
difficulty with investigating impacting systems is in finding instant of the
next impact what typically involves solving a nonlinear equation. However,
the problem can be simplified in the case of a bouncing ball dynamics
assuming a special motion of the limiter. Bouncing ball models have been
extensively studied, see \cite{Luo2009}\ and references therein. As a
motivation that inspired this work, we mention study of physics and
transport of granular matter \cite{Mehta1994}. A similar model has been also
used to describe the motion of railway bogies \cite{Knudsen1992}.

Recently, we have considered several models of motion of a material point in
a gravitational field colliding with a limiter moving with piecewise
constant velocity \cite{AOBR2008,AOBR2009a,AOBR2009b,AOBR2009c,AOBR2010a}.
Moreover, we have proposed more realistic yet still simple models
approximating sinusoidal motion of the table as exactly as possible but
still preserving possibility of analytical compuations \cite{AOBR2010b}. In
the present work we study the model in which displacement of the table is a
quadratic and periodic function of time.

The paper is organized as follows. In Section 2 a one dimensional dynamics
of a ball moving in a gravitational field and colliding with a table is
reviewed and the corresponding Poincare map is constructed. A bifurcation
diagram is computed for displacement of the table assumed as quadratic and
periodic function of time. In the next Section dynamical modes shown in the
bifurcation diagram such as fixed points, $2$ - cycles and chaotic bands are
studied analytically and numerically. It is shown that chaotic bands appear
due to homoclinic structures created from unstable $2$ - cycles in a
corner-type bifurcation. We summarize our results in Section 4.

\section{Bouncing ball: a simple motion of the table}

We consider a motion of a small ball moving vertically in a gravitational
field and colliding with a moving table, representing unilateral
constraints. The ball is treated as a material point while the limiter's
mass is assumed so large that its motion is not affected at impacts. A
motion of the ball between impacts is described by the Newton's law of
motion:%
\begin{equation}
m\ddot{x}=-mg,  \label{point motion}
\end{equation}%
where $\dot{x}=dx/dt$ and motion of the limiter is:%
\begin{equation}
y=y\left( t\right) ,  \label{limiter motion}
\end{equation}%
with a known function $y$. We shall also assume that $y$ is a continuous
function of time. Impacts are modeled as follows: 
\begin{eqnarray}
x\left( \tau _{i}\right) &=&y\left( \tau _{i}\right) ,  \label{position} \\
\dot{x}\left( \tau _{i}^{+}\right) -\dot{y}\left( \tau _{i}\right)
&=&-R\left( \dot{x}\left( \tau _{i}^{-}\right) -\dot{y}\left( \tau
_{i}\right) \right) ,  \label{velocity}
\end{eqnarray}%
where duration of an impact is neglected with respect to time of motion
between impacts. In Eqs. (\ref{position}), (\ref{velocity}) $\tau _{i}$
stands for time of the $i$-th impact while $\dot{x}_{i}^{-}$, $\dot{x}%
_{i}^{+}$are left-sided and right-sided limits of $\dot{x}_{i}\left(
t\right) $ for $t\rightarrow \tau _{i}$, respectively, and $R$ is the
coefficient of restitution, $0\leq R<1$ \cite{Stronge2000}.

Solving Eq. (\ref{point motion}) and applying impact conditions (\ref%
{position}), (\ref{velocity}) we derive the Poincar\'{e} map \cite{AOBR2007}%
: 
\begin{subequations}
\label{TV}
\begin{eqnarray}
\gamma Y\left( T_{i+1}\right) &=&\gamma Y\left( T_{i}\right) -\Delta
_{i+1}^{2}+\Delta _{i+1}V_{i},  \label{T} \\
V_{i+1} &=&-RV_{i}+2R\Delta _{i+1}+\gamma \left( 1+R\right) \dot{Y}\left(
T_{i+1}\right) ,  \label{V}
\end{eqnarray}%
where $\Delta _{i+1}\equiv T_{i+1}-T_{i}$. The limiter's motion has been
typically assumed in form $Y_{s}(T)=\sin (T)$, cf. \cite{AOBR2009a} and
references therein. This choice leads to serious difficulties in solving the
first of Eqs.(\ref{TV}) for $T_{i+1}$, thus making analytical investigations
of dynamics hardly possible. Accordingly, we have decided to simplify the
limiter's periodic motion to make (\ref{T}) solvable.

In our previous papers we have assumed displacement of the table as
piecewise linear periodic function of time \cite{AOBR2009a, AOBR2009b,
AOBR2010a}. In our recent work preliminary results for function $Y(T)$
assumed as quadratic $Y_{q}$ and two cubic functions of time, $Y_{c_{1}}$\
and $Y_{c_{2}}$ have been obtained \cite{AOBR2010b}. In this work we study
dynamics for quadratic function of time $Y_{q}\left( T\right) $: 
\end{subequations}
\begin{equation}
Y_{q}\left( T\right) =\left\{ 
\begin{array}{l}
-16\hat{T}\left( \hat{T}-\frac{1}{2}\right) ,\quad 0\leq \hat{T}<\frac{1}{2}
\\ 
16\left( \hat{T}-\frac{1}{2}\right) \left( \hat{T}-1\right) ,\quad \frac{1}{2%
}\leq \hat{T}\leq 1%
\end{array}%
\right.  \label{Q}
\end{equation}%
with $\hat{T}=T-\left\lfloor T\right\rfloor $, where $\left\lfloor
x\right\rfloor $ is the floor function -- the largest integer less than or
equal to $x$, in more detail.

Since the period of motion of the limiter is equal to one, the map (\ref{TV}%
) is invariant under the translation $T_{i}\rightarrow T_{i}+1$.
Accordingly, all impact times $T_{i}$ can be reduced to the unit interval $%
\left[ 0,\ 1\right] $. The model consists thus of equations (\ref{TV}), (\ref%
{Q}) with control parameters $R$, $\gamma $.

In Fig. 1 below we show the bifurcation diagram with times of impacts
computed for growing $\gamma $ and $R=0.85$ (see also \cite{AOBR2010b} where
bifurcation diagram with velocities just after impacts against $\gamma $ was
shown).

\begin{figure}[h]
\begin{equation*}
\includegraphics[width= 8 cm]{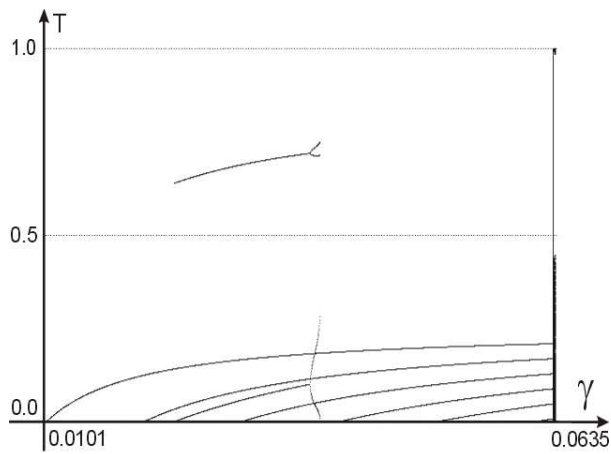}
\end{equation*}%
\caption{Bifurcation diagram. $R=0.85$, $\protect\gamma \in \left[ 0.0101,\
0.0635\right] $.}
\label{F1}
\end{figure}

It follows that dynamical system (\ref{TV}), (\ref{Q}) has several
attractors: six fixed points, one $2$ - cycle and, possibly, chaotic
attractor.

\section{Analytical results}

We shall investigate now fixed points, $2$ - cycle and chaotic bands shown
in Fig. 1, combining analytical and numerical approach.

\subsection{Fixed points}

We shall first study periodic solutions with one impact per $k$ periods.
Such states have to fulfill the following conditions:%
\begin{equation}
V_{n+1}=V_{n}\equiv V_{\ast }^{\left( k/1\right) },\ T_{n+1}=T_{n}+k\equiv
T_{\ast }^{\left( k/1\right) }+k\qquad \left( k=1,2,\ldots \right) ,
\label{condk1a}
\end{equation}%
where:%
\begin{equation}
T_{\ast }^{\left( k/1\right) }\in \left( 0,\ 1\right) ,\ V_{\ast }^{\left(
k/1\right) }>\gamma \dot{Y}_{q}\left( T\right) .  \label{condk1b}
\end{equation}

Substituting these conditions into (\ref{TV}), (\ref{Q}) we obtain two sets
of fixed points:%
\begin{eqnarray}
0 &\leq &T_{\ast }^{\left( s\right) }=\tfrac{1}{4}-\tfrac{k\left( 1-R\right) 
}{32\gamma \left( 1+R\right) }\leq \tfrac{1}{2}  \label{FixStA} \\
V_{\ast } &=&k  \notag
\end{eqnarray}%
where the impact occurs in time interval $T_{\ast }^{\left( s\right) }\in
\left( 0,\ \frac{1}{2}\right) $ and

\begin{eqnarray}
\tfrac{1}{2} &\leq &T_{\ast }^{\left( u\right) }=\tfrac{3}{4}+\tfrac{k\left(
1-R\right) }{32\gamma \left( 1+R\right) }\leq 1  \label{FixUStA} \\
V_{\ast } &=&k  \notag
\end{eqnarray}%
with impacts taking place in time interval $T_{\ast }^{\left( u\right) }\in
\left( \frac{1}{2},\ 1\right) $.

Solutions (\ref{FixStA}) fulfill physical requirements and are stable in the
following interval of $\gamma $:%
\begin{equation}
k\tfrac{1-R}{8\left( 1+R\right) }\leq \gamma \leq \tfrac{1+R^{2}}{8\left(
1+R\right) ^{2}},  \label{exst1}
\end{equation}%
where lower bound is a consequence of $T_{\ast }^{\left( s\right) }\geq 0$
while the upper bound follows from the condition that eigenvalues $\lambda $
of the stability matrix obey $\left\vert \lambda \right\vert <1$. On the
other hand, solutions (\ref{FixUStA}) are always unstable and are physical
for:%
\begin{equation}
\gamma \geq k\tfrac{1-R}{8\left( 1+R\right) },  \label{exust1}
\end{equation}%
what is equivalent to the condition $T_{\ast }^{\left( u\right) }\leq 1$.

\subsection{$2$ - cycle}

It follows from the bifurcation diagram, Fig. 1, that there exists a stable $%
2$ - cycle with time of first impact $T_{\ast 1}\in \left( 0,\ \frac{1}{2}%
\right) $ and time of the second impact $T_{\ast 2}\in \left( \frac{1}{2},\
1\right) $. Such periodic solution must fulfill the following equations
which are easily obtained from Eqs. (\ref{TV}), (\ref{Q}):%
\begin{equation}
\left\{ 
\begin{array}{l}
16\gamma \left( T_{\ast 2}-\tfrac{1}{2}\right) \left( T_{\ast 2}-1\right) =
\\ 
-16\gamma T_{\ast 1}\left( T_{\ast 1}-\tfrac{1}{2}\right) -\left( T_{\ast
2}-T_{\ast 1}\right) ^{2}+\left( T_{\ast 2}-T_{\ast 1}\right) V_{\ast 1} \\ 
V_{\ast 2}=-RV_{\ast 1}+2R\left( T_{\ast 2}-T_{\ast 1}\right) +\gamma \left(
1+R\right) \left( 32T_{\ast 2}-24\right)  \\ 
-16\gamma \left( T_{\ast 3}-1\right) \left( \left( T_{\ast 3}-1\right) -%
\tfrac{1}{2}\right) = \\ 
16\gamma \left( T_{\ast 2}-\tfrac{1}{2}\right) \left( T_{\ast 2}-1\right)
-\left( T_{\ast 3}-T_{\ast 2}\right) ^{2}+\left( T_{\ast 3}-T_{\ast
2}\right) V_{\ast 2} \\ 
V_{\ast 3}=-RV_{\ast 2}+2R\left( T_{\ast 3}-T_{\ast 2}\right) +\gamma \left(
1+R\right) \left( 8-32\left( T_{\ast 3}-1\right) \right)  \\ 
T_{\ast 3}=T_{\ast 1}+1 \\ 
V_{\ast 3}=V_{\ast 1}%
\end{array}%
\right. .  \label{2cycle}
\end{equation}

Eliminating variables we arrive at equation for time of the first impact
only:%
\begin{equation}
C_{4}x^{4}+C_{3}x^{3}+C_{2}x^{2}+C_{1}x+C_{0}=0,  \label{T1}
\end{equation}%
where $x\equiv T_{1\ast }$ and%
\begin{equation}
\left\{ 
\begin{array}{l}
C_{0}=\left( \left( 8\gamma +1\right) R^{2}-8\gamma +1\right) \tfrac{\left(
24\gamma +128\gamma ^{2}+1\right) R^{3}+\left( 384\gamma ^{2}+8\gamma
-1\right) R^{2}+\left( 384\gamma ^{2}-24\gamma +1\right) R+128\gamma
^{2}-8\gamma -1}{\left( R+1\right) ^{3}\left( R-1\right) ^{2}} \\ 
C_{1}=-64\gamma \tfrac{\left( 24\gamma +128\gamma ^{2}+1\right)
R^{4}-2R^{3}+\left( -512\gamma ^{2}+2+32\gamma \right) R^{2}+\left(
-2-512\gamma ^{2}+64\gamma \right) R-128\gamma ^{2}+8\gamma +1}{\left(
R-1\right) ^{2}\left( R+1\right) ^{2}} \\ 
C_{2}=-2048\gamma ^{2}\tfrac{-R^{3}+\left( -1+48\gamma \right) R^{2}+\left(
-5+48\gamma \right) R+1}{\left( R-1\right) ^{2}\left( R+1\right) } \\ 
C_{3}=4096\gamma ^{2}\left( -1+16\gamma \right) \tfrac{R^{2}+2R-1}{\left(
R-1\right) ^{2}} \\ 
C_{4}=-4096\gamma ^{2}\left( -1+16\gamma \right) \tfrac{R+1}{R-1}%
\end{array}%
\right. .  \label{C01234}
\end{equation}

It follows from the bifurcation diagram that the $2$ - cycle is born when $%
T_{1}=x=0$. This in turn occurs when $C_{0}=0$. Equation $C_{0}=0$ has three
roots:%
\begin{eqnarray}
\gamma _{1} &=&\tfrac{1}{8}\tfrac{1+R^{2}}{1-R^{2}},  \notag \\
\gamma _{2} &=&\tfrac{1}{32\left( R+1\right) }\tfrac{-3R^{2}+2R+1-\sqrt{%
R^{4}-12R^{3}-2R^{2}+4R+9}}{R+1},  \label{C0} \\
\gamma _{3} &=&\tfrac{1}{32\left( R+1\right) }\tfrac{-3R^{2}+2R+1+\sqrt{%
R^{4}-12R^{3}-2R^{2}+4R+9}}{R+1}.  \notag
\end{eqnarray}%
Testing Eqs. (\ref{C0}) against numerical computations we find out that the
stable $2$ - cycle is born at $\gamma =\gamma _{3}$. For example, for $%
R=0.85 $ we have $\gamma _{3}=0.023\,367\,4$, cf. Fig 1.

\subsection{Chaotic bands and homoclinic structure}

Magnification of the bifurcation diagram near the origin of chaotic bands is
shown in Fig. 2 below.

\begin{figure}[h]
\begin{equation*}
\includegraphics[width= 8 cm]{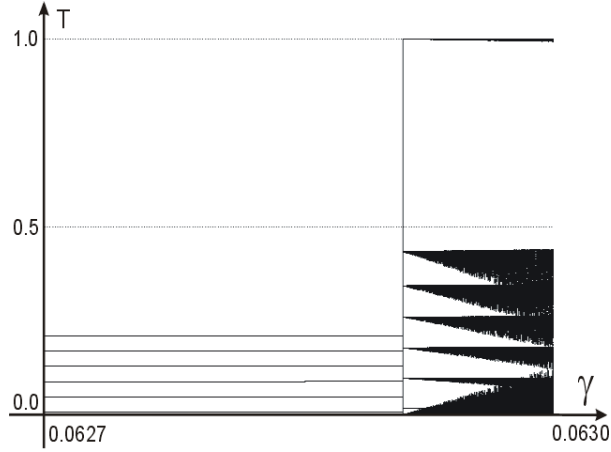}
\end{equation*}%
\caption{Bifurcation diagram. Six chaotic bands, $R=0.85$, $\protect\gamma %
\in \left[ 0.0627,\ 0.0630\right] $.}
\label{F2}
\end{figure}

There are six chaotic bands (and six basins of attraction) above the
critical point $\gamma _{cr}$. The first band which appears for appropriate
initial conditions is shown in Fig. 3. Each band consists of two subbands only 
since due to cyclic periodic conditions points $T=0$, $T=1$ are identified. 

\begin{figure}[h]
\begin{equation*}
\includegraphics[width= 8 cm]{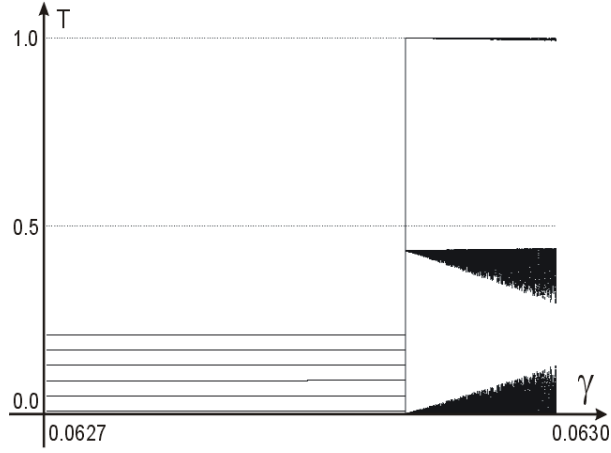}
\end{equation*}%
\caption{Bifurcation diagram. First chaotic band, $R=0.85$, $\protect\gamma %
\in \left[ 0.0627,\ 0.0630 \right] $.}
\label{F3}
\end{figure}

We note that there is a switch of stability - fixed points become unstable
precisely at $\gamma =\gamma _{cr}$ when chaotic bands appear. Then at $%
\gamma =\gamma _{cr}$ a homoclinic trajectory with $T=1^{-}$ is probably
created - this is suggested by presence of clusters of points near $T=0$, $%
T=1$. Computer simulations show that near $\gamma _{cr}$ there are six
unstable $2$ - cycles. For $\gamma <\gamma _{cr}$ impacts occur in the
following time intervals $T_{\ast 1}\in \left( 0,\ \frac{1}{2}\right) $, $%
T_{\ast 2}\in \left( 1,\ 1\frac{1}{2}\right) $,\ $T_{\ast 3}\in \left( 2+m,\
2+m+\frac{1}{2}\right) $, $m=0, 1, 2,\ldots $, and $T_{\ast 3}=T_{\ast 1} %
\mod \ 1$ where true impact times without cyclic conditions are shown. Then
at $\gamma =\gamma _{cr}$ a corner event occurs, i.e. $T_{\ast 2}=1$.
Indeed, at $T_{\ast 2}=1$ acceleration of the table is discontinuous. Then,
for $\gamma >\gamma _{cr}$ another six unstable $2$ - cycles are created
with impact times $T_{\ast 1}\in \left( 0,\ \frac{1}{2}\right) $, $T_{\ast
2}\in \left( \frac{1}{2},\ 1\right) $, $T_{\ast 3}\in \left( 2+m,\ 2+m+\frac{%
1}{2}\right) $, $m=0.1.2.\ldots $, and $T_{\ast 3}=T_{\ast 1} \mod 
\ 1$.

It is now possible to write down equations, suggested by numerical
computations, for the unstable $2$ - cycles for $\gamma \geq \gamma _{cr}$:

\begin{equation}
\left\{ 
\begin{array}{c}
16\gamma \left( y_{m}-\frac{1}{2}\right) \left( y_{m}-1\right) =-16\gamma
x\left( x-\frac{1}{2}\right) -\left( y-x\right) ^{2}+\left( y-x\right) u \\ 
v=-Ru+2R\left( y-x\right) +\gamma \left( 1+R\right) \left( 32y_{m}-24\right)
\\ 
-16\gamma z_{n}\left( z_{n}-\frac{1}{2}\right) =16\gamma \left( y_{m}-\frac{1%
}{2}\right) \left( y_{m}-1\right) -\left( z-y_{m}\right) ^{2}+\left(
z-y_{m}\right) v \\ 
w=-Rv+2R\left( z-y_{m}\right) +\gamma \left( 1+R\right) \left(
8-32z_{n}\right) \\ 
z_{n}=x \\ 
w=u%
\end{array}%
\right.  \label{U2cycles}
\end{equation}%
where%
\begin{eqnarray}
x &=&T_{\ast 1},\ y=T_{\ast 2},\ z=T_{\ast 3},\   \notag \\
y_{m} &=&T_{\ast 2}-m,\ z_{n}=T_{\ast 3}-n,  \label{defs} \\
u &=&V_{\ast 1},\ v=V_{\ast 2},\ w=V_{\ast 3},\   \notag
\end{eqnarray}%
with integer $m$, $n$ where numerical computations suggest that $n=m+2$.

We are going to solve equations (\ref{U2cycles}), (\ref{defs}) at $\gamma
=\gamma _{cr}$ and this means that we have to put $y_{m}=1$. It follows that
there are six unstable $2$ - cycles in question which are obtained for $%
m=0,1,\ldots ,5$ and $n=m+2$, solutions for larger $m$'s\ being nonphysical.
Solving these equations we get:%
\begin{eqnarray}
\gamma _{cr} &=&\tfrac{1+R^{2}}{8\left( 1+R\right) ^{2}},\quad n=m+2,
\label{gamma} \\
T_{\ast 1,\ m}^{\left( cr\right) } &\equiv &x=\tfrac{2R^{2}-6-4m+2\sqrt{%
\left( 3+5m+2m^{2}\right) R^{4}+1+3m+2m^{2}}}{2\left( R^{2}-1\right) }-m-2,
\label{T1sol}
\end{eqnarray}%
and we do not show more complicated expressions for $T_{\ast 2}^{\left(
cr\right) }$,$\ V_{\ast 1}^{\left( cr\right) }$,$\ V_{\ast 2}^{\left(
cr\right) }$.

It follows from Eq. (\ref{T1sol}) that for $R=0.85$ there are only six
acceptable solutions with $T_{\ast 1}>0$ corresponding to six chaotic bands
in Fig. 2. In the Table 1 impact times and the corresponding velocities just
after the impact, computed from Eqs. (\ref{U2cycles}), (\ref{gamma}), (\ref%
{T1sol}), are listed for $R=0.85$ and $m=0,1,\ldots ,5$:%
\begin{eqnarray*}
&&\text{Table 1} \\
&&%
\begin{tabular}{|l||l|l|l|l|l|l|}
\hline
$m$ & $0$ & $1$ & $2$ & $3$ & $4$ & $5$ \\ \hline
$T_{\ast 1,\ m}^{\left( cr\right) }$ & $0.\,434\,7$ & $0.\,343\,6$ & $%
0.\,260\,2$ & $0.\,178\,4$ & $0.\,097\,4$ & $0.\,016\,7$ \\ \hline
$V_{\ast 1,\ m}^{\left( cr\right) }$ & $0.\,514\,8$ & $1.\,623\,7$ & $%
2.\,716\,9$ & $3.\,806\,5$ & $4.\,894\,6$ & $5.\,981\,9$ \\ \hline
$T_{\ast 2,\ m}^{\left( cr\right) }$ & $1^{-}$ & $1^{-}$ & $1^{-}$ & $1^{-}$
& $1^{-}$ & $1^{-}$ \\ \hline
$V_{\ast 2,\ m}^{\left( cr\right) }$ & $1.\,454\,6$ & $2.\,366\,7$ & $%
3.\,279\,4$ & $4.\,192\,3$ & $5.\,105\,1$ & $6.\,018\,0$ \\ \hline
\end{tabular}%
\end{eqnarray*}

Critical $2$ - cycle $m=0$, $n=2$ $\left( R=0.85\right) $\ is shown below.

\newpage

\begin{figure}[h]
\begin{equation*}
\includegraphics[width= 8 cm]{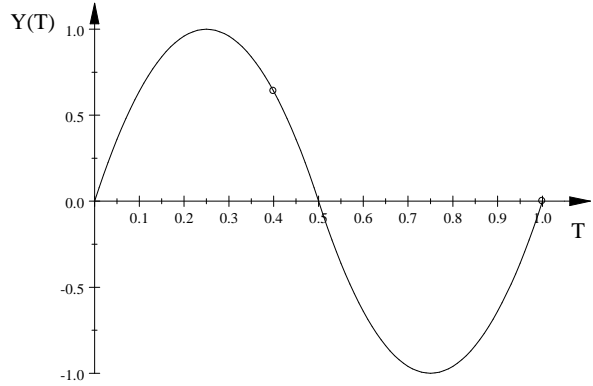}
\end{equation*}%
\caption{Critical $2$ - cycle. Position of the $2$ - cycle is denoted by
open circles.}
\label{F4}
\end{figure}

In Figs. 5, 6 the second chaotic band and first two chaotic bands are shown.
The bifurcation diagrams for $\gamma >\gamma _{cr}=0.06291\ldots $ were
computed for initial conditions shown in the Table 1.

Sharp edges of chaotic bands are given within good approximation by $T_{\ast
1,\ m}^{\left( cr\right) }$ and also $T=0,1$. It seems that the homoclinic
structure exists for all values of $\gamma >\gamma _{cr}$ shown in the
Figures.

\bigskip

\begin{figure}[h]
\begin{equation*}
\includegraphics[width= 8 cm]{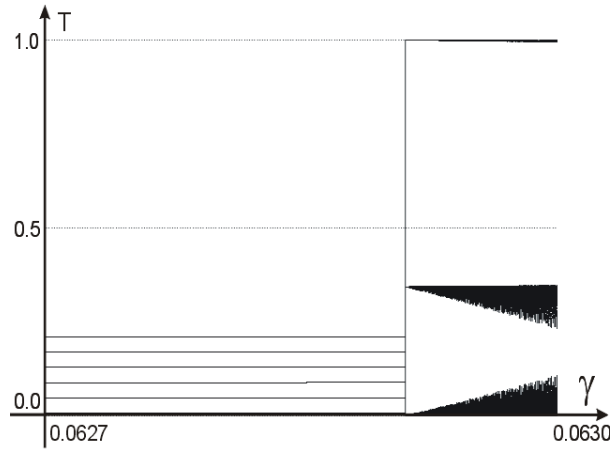}
\end{equation*}%
\caption{Bifurcation diagram. Second chaotic band, $R=0.85$, $\protect\gamma %
\in \left[ 0.0627,\ 0.0630\right] $.}
\label{F5}
\end{figure}

\newpage

\begin{figure}[h]
\begin{equation*}
\includegraphics[width= 8 cm]{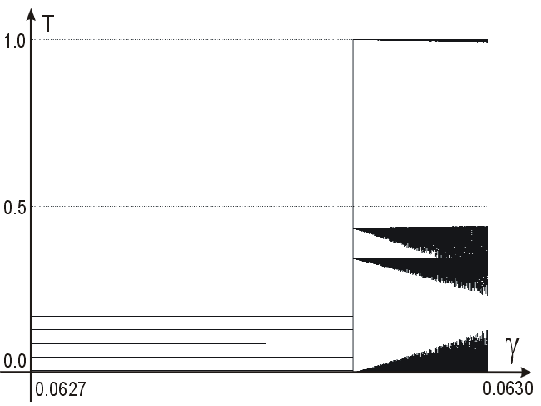}
\end{equation*}%
\caption{Bifurcation diagram. First two chaotic bands, $R=0.85$, $\protect%
\gamma \in \left[ 0.0627,\ 0.0630\right] $.}
\label{F6}
\end{figure}

We have also computed Lyapunov exponents, cf. the Table 2. 
\begin{eqnarray*}
&&\text{Table 2} \\
&&%
\begin{tabular}{|l||l|l|l|l|l|l|}
\hline
$m$ & $0$ & $1$ & $2$ & $3$ & $4$ & $5$ \\ \hline
$\lambda _{m}$ & $0.2$ & $1.2$ & $1.6$ & $1.85$ & $2.0$ & $2.1$ \\ \hline
\end{tabular}%
\end{eqnarray*}

\section{Discussion and closing remarks}

We have studied dynamics of a material point moving vertically in a
gravitational field and colliding with a limiter. Displacement of the
limiter has been assumed as quadratic function of time (\ref{Q}). Due to the
simplicity of the problem it was possible to investigate the dynamics
analytically with some support from numerical computations. Firstly, fixed
points were found and their stability was determined. Secondly, equations
for a stable $2$ - cycle were found and simplified, cf. Eqs. (\ref{T1}), (%
\ref{C01234}). From these equations analytical condition for birth of the $2$
- cycle was found (cf. $\gamma =\gamma _{3}$ in Eq.(\ref{C0})). Finally, a
transition to chaotic dynamics was described in analytical terms. It was
shown that six stable chaotic bands appear from six unstable $2$ - cycles.
Equations for these $2$ - cycles were found and solved to yield critical
value of $\gamma $, Eq. (\ref{gamma}), and impact times and the
corresponding velocities at $\gamma =\gamma _{cr}$, see Eq. (\ref{T1sol})
where $T_{\ast 1,\ m}^{\left( cr\right) }$ was given. Approximation to the
band edges was also found.

We have demonstrated, combining analytical and numerical approach, that at
the transition point $\gamma =\gamma _{cr}$ unstable $2$ - cycles give rise
to homoclinic structures which lead to chaotic behaviour. This transition is
a corner-type bifurcation similar to that found in a bouncing ball model
with piecewise linear velocity \cite{AOBR2010a}. In our future work we shall
study models with displacement of the table described by a cubic functions
of time \cite{AOBR2010b}.

\end{document}